\documentclass[a4paper,11pt]{article}

\newcommand{\be}{\begin{equation}}
\newcommand{\ee}{\end{equation}}
\newcommand{\la}{\langle}
\newcommand{\ra}{\rangle}
\newcommand{\mG}{\mathcal{G}}

\title{ERD\'ELYI--KOBER FRACTIONAL DIFFUSION}

\author{Gianni PAGNINI\\
CRS4, 09010 Pula (CA), Italy,}

\begin{document}
\maketitle

\begin{abstract}
The aim of this Short Note is to highlight that the 
{\it generalized grey Brownian motion} (ggBm) is an anomalous diffusion process driven by
a fractional integral equation in the sense of Erd\'elyi--Kober, and for this
reason here it is proposed to call such family of diffusive processes as {\it Erd\'elyi--Kober fractional diffusion}.
The ggBm is a parametric class of stochastic processes that provides models for both fast and slow anomalous diffusion.
This class is made up of self-similar processes with stationary increments and 
it depends on two real parameters: $0 < \alpha \le 2$ and $0 < \beta \le 1$. 
It includes the fractional Brownian motion when $0 < \alpha \le 2$ and $\beta=1$,
the time-fractional diffusion stochastic processes when $0 < \alpha=\beta <1$,
and the standard Brownian motion when $\alpha=\beta=1$. 
In the ggBm framework, the Mainardi function  
emerges as a natural generalization of the Gaussian distribution recovering the same key role of the Gaussian density for
the standard and the fractional Brownian motion. 
\end{abstract}

 \medskip


 \smallskip


\vspace{2truecm}

Diffusion in disordered media is commonly named {\it anomalous}, 
to mark a difference with the classical diffusion processes where the probability density function $pdf$ 
to find a particle in the place $x$ at the time $t$ is Normal, i.e., Gaussian, 
and for this it is called {\it Normal} diffusion, or {\it Gaussian} diffusion.
Moreover, beside the non-Gaussian particle $pdf$, the anomalously is embodied even by  
the non-linear growing in time of the variance of the particle spreading \cite{klafter_etal-pw-2005}. 
The process of anomalous diffusion is referred to as {\it fast diffusion},
when the variance of the particle spreading grows with a power law with exponent greater than $1$, 
and it is referred to as {\it slow diffusion},
when such exponent is lower than $1$.

In the case of anomalous diffusion, the classical local flux-gradient relationship  
does not hold and it is necessary to determine a non-local relationship. 
It is well-known that a useful mathematical tool for physical investigation and description of non-local and 
anomalous diffusion is Fractional Calculus,
which is that branch of mathematical analysis dealing with pseudo-differential operators interpreted as 
integrals and derivatives of non-integer order \cite{kiryakova-1994,podlubny-1999}. 

Non-locality can be designated in time (time-fractional diffusion) or in space (space-fractional diffusion), 
as well as both in space and time (space-time fractional diffusion equation) \cite{mainardi_etal-fcaa-2001}. 
Generally, when the fractional differentiation is considered for the time,
then the fractional derivative operator is assumed to be in the Caputo or in the Riemann--Liouville sense,
when the fractional differentiation is considered for the space, then the fractional derivative operator is
assumed to be in the Riesz--Feller sense.

Recently, the extension of fractional differential equations to distributed-order fractional differential equations 
has permitted to describe also processes whose scaling law changes in time, 
see e.g. \cite{mainardi_etal-jcam-2007,saxena_etal-pa-2011,sokolov_etal-appb-2004}.

Furthermore, under the physical point of view,
when there is no separation of timescale between the microscopic and the macroscopic level of the process 
the randomness of the microscopic level is transmitted to the macroscopic level and 
the correct description of the macroscopic dynamics has to be in terms of the Fractional Calculus \cite{grigolini_etal-pre-1999}.
Moreover, fractional integro/differential equations are related to phenomena with fractal properties \cite{rocco_etal-pa-1999}. 

A fractional differential approach has been successfully used for modelling in several different disciplines
as for example statistical physics \cite{metzler_etal-jpa-2004}, neuroscience \cite{lundstrom_etal-nn-2008}, 
economy \cite{scalas-pa-2006}, control theory \cite{vinagre_etal-fcaa-2000} and combustion science \cite{pagnini-fcaa-2011}.
Further applications of the fractional approach are recently introduced and discussed by 
J. A. Tenreiro Machado \cite{tenreiro-machado-fcaa-2011}.

Normal diffusion, or Gaussian diffusion, is a Markovian stochastic process driven by 
the classical parabolic equation
\be
\frac{\partial P}{\partial t}=
\frac{\partial^2 P}{\partial x^2} \,,
\quad x \in \mathcal{R} \,, \quad t \in \mathcal{R}^+_0 \,,
\label{brownian}
\ee
with initial condition $P(x,0)=P_0(x)$. 
The fundamental solution of (\ref{brownian}), which is named also Green function, 
and corresponding to the case with initial condition $P(x,0)=P_0(x)=\delta(x)$,
is the Gaussian density
\be
f(x,t)=\frac{1}{\sqrt{4 \pi t}} \exp\left\{-\frac{x^2}{4 t}\right\} \,,
\label{gaussian}
\ee
whose variance grows linearly in time,
i.e., $\displaystyle \la x^2 \ra =\int_{-\infty}^{+\infty} x^2 f(x,t) \, dx = 2 \, t$.
The density function $P(x,t)$ with general initial condition $P(x,0)=P_0(x)$ is related to the fundamental solution $f(x,t)$
by the following convolution integral
\be
P(x,t)=\int_{-\infty}^{+\infty} f(\xi,t)P_0(x-\xi) \, d\xi \,.
\ee

In order to generalize the classical Markovian setting to Non-Markovian cases,
the following integral equation has been introduced by Mura, Taqqu and Mainardi
\cite{mura_etal-pa-2008}
\be
P(x,t)=P_0(x)+\int_0^t \frac{\partial g(s)}{\partial s} K[g(t)-g(s)]
\frac{\partial^2 P(x,s)}{\partial x^2} \, ds \,,
\label{mtm-equation}
\ee
where $K(t)$ is a memory kernel and $g(t)$, with $g(0)=0$,
is a smooth and increasing function describing a time stretching.
The Green function of (\ref{mtm-equation}) $\mG(x,t)$,
which is the marginal one-point one-time $pdf$ of the non-Markovian diffusion process,
turns out to be
\be
\mG(x,t)=\int_0^\infty f(x,\tau) h(\tau,g(t)) \, d\tau \,,
\label{green-mtm}
\ee
where $f(x,t)$ is the Gaussian density (\ref{gaussian}) that is the fundamental solution of the Markovian diffusion process,
i.e., $K(t)=\delta(t)$,
and $h(\tau,t)$ is the fundamental solution of the so-called {\it non-Markovian forward drift equation}
\be
u(\tau,t)=u_0(\tau)-\int_0^t K(t-s) \frac{\partial u(\tau,s)}{\partial \tau} \, ds \,,
\quad \tau \,,t \in \mathcal{R}^+_0 \,,
\label{h-equation}
\ee
where $u_0(\tau)=u(\tau,0)$.

When the kernel and the time-stretching functions are stated as 
\be
K(t)=\frac{t^{\beta -1}}{\Gamma(\beta)} \,, \quad
g(t)=t^{\alpha/\beta} \,, \quad 0 < \alpha \le 2 \,, \quad 0 < \beta \le 1 \,,
\label{specialcase}
\ee
Equation (\ref{mtm-equation}) becomes 
\be
P(x,t)=P_0(x) + \frac{1}{\Gamma(\beta)}\frac{\alpha}{\beta}\,
\int_0^t \tau^{\alpha/\beta-1}\, (t^{\alpha/\beta}-\tau^{\alpha/\beta})^{\beta-1}
\,\frac{\partial^2 P(x,\tau)}{\partial x^2} \, d\tau \,,
\label{fractionalkineticequation}
\ee
that was originally introduced by A. Mura in his PhD Thesis \cite{mura-phd-2008},
and later discussed by him and collaborators in a number of papers 
\cite{mainardi_etal-ijde-2010,mura_etal-itsf-2009,mura_etal-jpa-2008,mura_etal-pa-2008}.

It is well-known that it exists a relationship between the solutions of a certain class of 
integral equations that are used to model anomalous diffusion and stochastic processes.
In this respect, 
the density function $P(x,t)$ which solves (\ref{fractionalkineticequation}) is the marginal particle $pdf$,
i.e., the one-point one-time density function of particle dispersion, 
of the {\it generalized grey Brownian motion} (ggBm)
\cite{mura-phd-2008,mura_etal-itsf-2009,mura_etal-jpa-2008}.

The ggBm is a special class of H-sssi processes of order $H=\alpha$, or Hurst exponent $H=\alpha/2$,
where, according to a common terminology, H-sssi means H-self-similar-stationary-increments.
The ggBm provides non-Markovian stochastic models for anomalous diffusion, 
of both slow type $0 < \alpha < 1$ and fast type $1 < \alpha <2$.
The ggBm includes some well-known processes, so that it defines an interesting general theoretical framework. 
In fact, the fractional Brownian motion appears for $\beta=1$,
the grey Brownian motion, in the sense of W. R. Schneider \cite{schneider-1990,schneider-1992}, 
corresponds to the choice $0< \alpha=\beta <1$,
and finally the standard Brownian motion is recovered by setting $\alpha=\beta=1$.
It is worth noting to remark that only in the particular case of the Brownian motion the stochastic process is Markovian. 
Moreover, the ggBm is not an ergodic process \cite{mura_etal-jpa-2008}.

The integral in the non-Markovian kinetic equation (\ref{fractionalkineticequation})
can be expressed in terms of an Erd\'elyi--Kober fractional integral.
In fact, 
let $\mu$, $\eta$ and $\gamma$ be $\mu >0$, $\eta >0$ and $\gamma \in \mathcal{R}$,
the Erd\'elyi--Kober fractional integral operator $I^{\gamma,\mu}_\eta$,
for a sufficiently well-behaved function $\varphi(t)$, is defined as \cite[formula (1.1.17)]{kiryakova-1994}
\begin{eqnarray}
I^{\gamma,\mu}_{\eta} \, \varphi(t)
&=&\frac{t^{-\eta(\mu+\gamma)}}{\Gamma(\mu)} 
\int_0^t \tau^{\eta \gamma} \, (t^\eta-\tau^\eta)^{\mu-1} \varphi(\tau) \, d(\tau^\eta) \,, \nonumber \\
&=&\frac{\eta}{\Gamma(\mu)} t^{-\eta(\mu+\gamma)}
\int_0^t \tau^{\eta(\gamma+1)-1} (t^\eta-\tau^\eta)^{\mu-1} \varphi(\tau) \, d\tau \,,
\label{EK}
\end{eqnarray}
hence Equation (\ref{fractionalkineticequation}) can be re-written as
\be
P(x,t)=P_0(x)+t^\alpha \, \left[I^{0,\beta}_{\alpha/\beta} \, \frac{\partial^2 P}{\partial x^2}\right] \,.
\label{EK-equation}
\ee

The integral operator $I^{\gamma,\mu}_\eta$ was introduced by I. N. Sneddon (see for example \cite{sneddon-1966,sneddon-lnm-1975,sneddon-1979}) 
who studied its basic properties and emphasized its useful applications to the generalized axially symmetric potential 
theory (GASPT) and other physical problems (say in electrostatics, elasticity, etc). 
When $\eta=1$, one obtains the operators of fractional integration as originally introduced by 
H. Kober \cite{kober-qjmo-1940} and A. Erd\'elyi \cite{erdelyi-qjmo-1940} and, when $\eta=2$, 
those introduced by I. N. Sneddon \cite{sneddon-1966,sneddon-lnm-1975,sneddon-1979}.
In the special case $\gamma=0$ and $\eta=1$, the Erd\'elyi--Kober fractional integral operator (\ref{EK})
and the Riemann--Liouville fractional integral of order $\mu$, here noted by $J^\mu$,
are related by the formula
\be
I^{0,\mu}_{1} \, \varphi(t)
=\frac{t^{-\mu}}{\Gamma(\mu)} \,
\int_0^t (t-\tau)^{\mu-1} \varphi(\tau) \, d\tau=t^{-\mu} J^\mu \, \varphi(t) \,.
\ee

The above remark about the relationship between the governing equation of the ggBm (\ref{fractionalkineticequation}) 
and the Erd\'elyi--Kober operator of Fractional Calculus (\ref{EK}) constitutes the aim of this Short Note.
The possibility to re-write Equation (\ref{fractionalkineticequation}) as (\ref{EK-equation}) 
was briefly noted by the Author in \cite{pagnini-epjst-2011}.
However, here this correspondence between the ggBm and the Erd\'elyi--Kober fractional integral operator is stressed and, 
since the ggBm serves as a stochastic model for the anomalous diffusion, 
this leads to define the family of diffusive processes governed by the ggBm as {\it Erd\'elyi--Kober fractional diffusion}.

In order to establish the diffusion-type equation corresponding to (\ref{fractionalkineticequation}),
the Erd\'elyi-Kober fractional differential operator is here introduced.
Let $n-1 < \mu \le n$, $n \in \mathcal{N}$,
the Erd\'elyi--Kober fractional derivative is defined as \cite[formula (1.5.19)]{kiryakova-1994}
\be
D^{\gamma,\mu}_\eta \, \varphi(t)=\prod_{j=1}^n \left(\gamma+j+\frac{1}{\eta}t\frac{d}{dt}\right)(I^{\gamma+\mu,n-\mu}_\eta \, \varphi(t)) \,.
\label{EK-derivative}
\ee
The Riemann--Liouville fractional derivative of order $\mu$, $m-1 < \mu \le m$, $m \in \mathcal{N}$ 
is defined as $\displaystyle D^{\mu}_{RL} \, \varphi(t)=\frac{d^m}{dt^m} J^{m-\mu} \, \varphi(t)$,
and it emerges that the Erd\'elyi--Kober and the Riemann--Liouville fractional derivatives are related through the formula
\be
D^{-\mu,\mu}_1 \, \varphi(t) = t^{\mu} D^{\mu}_{RL} \, \varphi(t) \,.
\label{EK-RL}
\ee
A further important property of the Erd\'elyi--Kober fractional derivative is the reduction to the
identity operator when $\mu=0$, i.e., 
\be
D^{\gamma,0}_\eta \varphi(t)=\varphi(t) \,.
\label{EK-derivative-identity}
\ee

Recently, the notions of Erd\'elyi--Kober fractional integrals and derivatives have been further 
extended by Yu. Luchko \cite{luchko-fcaa-2004} and Yu. Luchko \& J. Trujillo \cite{luchko_etal-fcaa-2007}. 

Equation (\ref{EK-equation}) in diffusive form is obtained by deriving in time both sides
and it results
\begin{eqnarray}
\frac{\partial P}{\partial t} 
&=& \alpha \, t^{\alpha-1} I^{0,\beta}_{\alpha/\beta} \frac{\partial^2 P}{\partial x^2} 
+ t^\alpha \frac{\partial}{\partial t}\left(I^{0,\beta}_{\alpha/\beta} \frac{\partial^2 P}{\partial x^2}\right) \,, \nonumber \\ 
&=& t^{\alpha-1} \left[\alpha + t \frac{\partial}{\partial t}\right] \left(I^{0,\beta}_{\alpha/\beta} \frac{\partial^2 P}{\partial x^2}\right) \,,
\end{eqnarray}
that can be recast as
\be
\frac{\partial P}{\partial t} 
= \frac{\alpha}{\beta} t^{\alpha-1} \left[(\beta-1) + 1 + \frac{\beta}{\alpha} t \frac{\partial}{\partial t}\right] 
\left(I^{0,\beta}_{\alpha/\beta} \frac{\partial^2 P}{\partial x^2}\right) \,,
\ee
and finally, by using (\ref{EK-derivative}), 
\be
\frac{\partial P}{\partial t} 
= \frac{\alpha}{\beta} t^{\alpha-1} \, 
D^{\beta-1,1-\beta}_{\alpha/\beta} \, \frac{\partial^2 P}{\partial x^2} \,.
\label{EK-diffequation}
\ee
A diffusion-type equation for the ggBm was previously proposed \cite{mainardi_etal-ijde-2010}
but adopting, with an abuse of notation, the Riemann--Liouville fractional differential operator with a stretched time variable.
Then, since the Erd\'elyi--Kober fractional differential operator is taken into account,
Equation (\ref{EK-diffequation}) follows to be the correct formulation.

The Green function corresponding to (\ref{EK-equation}, \ref{EK-diffequation}) 
is \cite{mura-phd-2008,mura_etal-itsf-2009,mura_etal-jpa-2008,mura_etal-pa-2008}
\be
\mG(x,t)= \frac{1}{2} \, \frac{1}{t^{\alpha/2}}
M_{\beta/2}\left(\frac{|x|}{t^{\alpha/2}}\right) \,,
\label{green-stretched-fractional}
\ee
where $M_\nu(z)$ is the $M$-Wright function, 
often referred to as Mainardi function in the literature devoted to fractional diffusion \cite{mainardi-2010,podlubny-1999},
and it is defined as \cite{mainardi-csf-1996}
\begin{eqnarray}
M_\nu(z) 
&=& \sum_{n=0}^\infty \frac{(-z)^n}{n! \, \Gamma[-\nu n + (1-\nu)]} \,, \nonumber \\
&=& \frac{1}{\pi} \sum_{n=1}^\infty \frac{(-z)^{n-1}}{(n-1)!} \, \Gamma(\nu n)\sin(\pi \nu n) \,,
\quad 0 < \nu < 1 \,,
\label{M-function}
\end{eqnarray}
see Reference \cite{gorenflo_etal-fcaa-1999,gorenflo_etal-jcam-2000,mainardi_etal-ijde-2010} for a review. 
Here it is reminded 
the noteworthy composition, or subordination-type, formula \cite{mainardi_etal-fcaa-2003}
\be
t^{-\nu} M_\nu\left(\frac{\xi}{t^\nu}\right)= t^{-\ell} \, \int_0^\infty
M_\lambda\left(\frac{\xi}{\tau^\lambda}\right) M_\ell\left(\frac{\tau}{t^\ell}\right) \, \frac{d\tau}{\tau^\lambda} \,,
\quad {\rm with} \quad \nu=\lambda \, \ell \,, 
\label{Mcomposition}
\ee
where $0 < \nu \,, \lambda \,, \ell < 1$ and $\xi \,, t \,, \tau \in \mathcal{R}_0^+$.
By using (\ref{Mcomposition}) and the special case $M_{1/2}(z)=(1/\sqrt{\pi})\exp(-z^2/4)$,
Green function (\ref{green-stretched-fractional}) can be expressed as 
\cite{mainardi_etal-ijde-2010,mura_etal-jpa-2008,mura_etal-pa-2008}
\begin{eqnarray}
\mG(x,t)
&=& \frac{1}{2} \, \frac{1}{t^{\alpha/2}}
M_{\beta/2}\left(\frac{|x|}{t^{\alpha/2}}\right) \,, \\
&=& \frac{1}{\sqrt{4t^\alpha}} \, \int_0^\infty M_{1/2}\left(\frac{|x|t^{-\alpha/2}}{\tau^{1/2}}\right) \, 
M_\beta(\tau) \, d\tau \,, \\
&=& \int_0^\infty \frac{1}{\sqrt{4 \pi \tau t^\alpha}} \exp\left\{-\frac{x^2}{4 \tau t^\alpha}\right\} \, 
M_\beta(\tau) \, d\tau \,,
\label{fBmdistributed}
\end{eqnarray}
so that, under the view point of statistical mechanics, the ggBm, 
or the {\it Erd\'elyi--Kober fractional diffusion}, 
emerges to be the superposition of processes with stretched Gaussian density 
$\displaystyle \frac{1}{\sqrt{4 \pi \tau t^\alpha}} \exp\left\{-\frac{x^2}{4 \tau t^\alpha}\right\}$,
i.e. fractional Brownian motions,
whose variance is $\la x^2 \ra=2 \tau t^\alpha$ where $\tau$ is a random coefficient distributed according to $M_\beta(\tau)$.

However, Equation (\ref{fBmdistributed}) can be further re-managed to exhibit a subordination type representation.
In fact, after the change of variable $t_*=\tau t^\alpha$, it follows that
\be
\mG(x,t)=
\int_0^\infty \frac{1}{\sqrt{4 \pi t_*}} \exp\left\{-\frac{x^2}{4 t_*}\right\} \, 
\frac{1}{t^\alpha} M_\beta\left(\frac{t_*}{t^\alpha}\right) \, d t_* \,,
\label{gorenflo}
\ee
which means that the random trajectory $x=x(t)$ can be obtained as a subordination process by $x=x(t)=y[t_*(t)]$,
where $t_*=t_*(t)$ is a positive random variable that evolves in the natural time $t$ and is referred to as operational time
\cite{gorenflo_etal-epjst-2011,gorenflo_etal-fdra-2011}.
The process $t_*=t_*(t)$ is the directing process that realizes in the ($t$, $t_*$)-plane
whose $pdf$ is $t^{-\alpha}M_\beta(t_*t^{-\alpha})$. Please note that the $pdf$ of the directing process belongs 
to the same family of the Green function $\mG(x,t)$ and they differ for the parameter pair.
The process $y=y(t_*)$ is the parent process that is a random trajectory in the ($t_*$, $y$)-plane with Gaussian $pdf$ evolving in
the operational time $t_*$.
Geometrically, identifying the spatial coordinates $y$ and $x$,
the subordination structure $x=x(t)=y[t_*(t)]$ is obtained by concatenation.

The marginal $pdf$ of the non-Markovian diffusion process ggBm emerges to be related to 
the Mainardi function $M_\nu$ and it describes both slow and fast anomalous diffusion.
In fact, the variance of Green function (\ref{green-stretched-fractional}) is
$\displaystyle \langle x^2 \rangle=\int_{-\infty}^{+\infty} x^2 \mG(x,t) \, dx =  (2/\Gamma(\beta+1)) \, t^\alpha$,
then the resulting process turns out to be self-similar with Hurst exponent $H=\alpha/2$
and the variance law is consistent with slow diffusion for  $0<\alpha <1$
and fast diffusion for $1<\alpha \le 2$.
However it is worth noting to be remarked also that a linear variance growing is possible, but with non-Gaussian $pdf$,
when $\beta \ne \alpha=1$, and a Gaussian $pdf$ with non-linear variance growing when $\beta=1$ and $\alpha \ne 1$.

It is straightforward to note that, by using formula (\ref{EK-RL}), 
evolution equation (\ref{fractionalkineticequation})
reduces to the time-fractional diffusion if $\alpha=\beta < 1$, i.e.,
\be
\frac{\partial P}{\partial t}= D^{1-\beta}_{RL} \, \frac{\partial^2 P}{\partial x^2} \,,
\quad {\rm with} \quad
\mG(x,t)=\frac{1}{2}\frac{1}{t^{\beta/2}} M_{\beta/2}\left(\frac{|x|}{t^{\beta/2}}\right) \,,
\label{green-time-frac}
\ee
and variance $\langle x^2 \rangle= (2/\Gamma(\beta+1)) \, t^\beta$,
and, by using formula (\ref{EK-derivative-identity}),
it reduces to the stretched Gaussian diffusion if $\alpha \ne 1$ and $\beta =1$, i.e.,
\begin{eqnarray}
\!\!\!\!
\frac{\partial P}{\partial t}= \alpha \, t^{\alpha-1} \, \frac{\partial^2 P}{\partial x^2} \,,
\quad {\rm with} \quad \mG(x,t) 
\!\! &=& \!\! \frac{1}{2}\frac{1}{t^{\alpha/2}} M_{1/2}\left(\frac{|x|}{t^{\alpha/2}}\right) \nonumber \\
\!\! &=& \!\! \frac{1}{\sqrt{4 \pi}}\frac{1}{t^{\alpha/2}} \exp\left\{-\frac{x^2}{4t^\alpha}\right\} \,,
\label{green-stretched-gauss}
\end{eqnarray}
and variance $\langle x^2 \rangle= 2 \, t^\alpha$,
and finally to the standard Gaussian diffusion if $\alpha=\beta=1$, i.e., (\ref{brownian}) with (\ref{gaussian})
and variance $\langle x^2 \rangle=2 \, t$.
The Green functions of these last two cases, i.e., ($\alpha \ne 1$, $\beta=1$) and ($\alpha=\beta=1$), 
follows by (\ref{fBmdistributed}) noting that $M_1(\tau)=\delta(\tau-1)$.

In general, even if the Green functions are interpreted as one-point $pdf$ evolving
in time, they cannot determine a {\it unique} (self-similar) stochastic process
because this requires the determination of any multi-point $pdf$.
But, what concerns the ggBm, since the increments are stationary, 
it emerges to be uniquely determined by its covariance structure \cite{mura_etal-itsf-2009,mura_etal-jpa-2008}.
Then, even if the ggBm is not Gaussian in general, it is a valuable example of a process defined only through its first and second moments, 
which indeed is a remarkable property of the Gaussian processes. 
Then the ggBm is a direct generalization of the Gaussian processes and, in the same way,
the Mainardi function $M_\nu$ is a generalization of the Gaussian function, 
and it emerges to be the marginal $pdf$ of non-Markovian diffusion processes
that describe both slow and fast anomalous diffusion.

To conclude, in this Short Note it is highlighted the relationship between the Erd\'elyi--Kober fractional operators
and the valuable family of stochastic processes generated by the ggBm, 
whose some remarkable properties are reported above, and the key role of the Mainardi function in this framework.
In fact, the particle $pdf$ of associated to the ggBm 
is the solution of a fractional integral equation (\ref{EK-equation}),
or analogously of a fractional diffusion equation (\ref{EK-diffequation}), in the Erd\'elyi--Kober sense
and this solution is a Mainardi function.
Since the governing equation of these processes is 
a fractional equation in the Erd\'elyi--Kober sense
it is proposed to called this family of diffusive processes as {\it Erd\'elyi--Kober fractional diffusion}.


\smallskip
\section*{Acknowledgements}
The author would like to thank Professor Francesco Mainardi for the advice and encouragement,
and Professor Rudolf Gorenflo and the journal Managing Editor Professor Virginia Kiryakova for the help to improve this Short Note.
The research is funded by Regione Autonoma della Sardegna 
(PO Sardegna FSE 2007--2013 sulla L.R. 7/2007 
``Promozione della ricerca scientifica e dell'innovazione tecnologica in Sardegna'').



\begin{thebibliography}{99}
\normalsize
\bibitem{erdelyi-qjmo-1940}
A. Erd\'elyi, 
On fractional integration and its applications to the theory of Hankel transforms. 
\emph{Quart. J. Math. Oxford} {\bf 11}, No 1 (1940), 293--303.

\bibitem{gorenflo_etal-fcaa-1999}
R. Gorenflo, Yu. Luchko, F. Mainardi,
Analytical properties and applications of the Wright function.
\emph{Fract. Calc. Appl. Anal.} {\bf  2}, No 4 (1999), 383--414.

\bibitem{gorenflo_etal-jcam-2000}
R. Gorenflo, Yu. Luchko, F. Mainardi,
Wright functions as scale-invariant solutions of the diffusion-wave equation.
{\it J. Comput. Appl. Math.} {\bf 118}, No 1-2 (2000), 175--191.

\bibitem{gorenflo_etal-epjst-2011}
R. Gorenflo, F. Mainardi,
Subordination pathways to fractional diffusion.
{\it Eur. Phys. J. Special Topics} {\bf 193}, (2011), 119--132.

\bibitem{gorenflo_etal-fdra-2011}
R. Gorenflo, F. Mainardi,
Parametric subordination in fractional diffusion processes.
In: {\it Fractional Dynamics. Recent Advances},
J. Klafter, S.C. Lim and R. Metzler (Editors),
World Scientific, Singapore (2011), Chapter 10, 227--261. 

\bibitem{grigolini_etal-pre-1999}
P. Grigolini, A. Rocco, B.J. West, 
Fractional calculus as a macroscopic manifestation of randomness. 
{\it Phys. Rev. E} {\bf 59}, No 3 (1999), 2603--2613.

\bibitem{kiryakova-1994}
V. Kiryakova,
{\it Generalized Fractional Calculus and Applications}.
Longman Scientific \& Technical, Harlow (1994).

\bibitem{klafter_etal-pw-2005}
J. Klafter, I.M. Sokolov,
Anomalous diffusion spreads its wings.
{\it Physics World}, August (2005), 29--32.

\bibitem{kober-qjmo-1940}
H. Kober, 
On a fractional integral and derivative. 
\emph{Quart. J. Math. Oxford} {\bf 11}, No 1 (1940), 193--211.

\bibitem{luchko-fcaa-2004}
Yu. Luchko, Operational rules for a mixed operator of the Erd\'elyi-Kober type. 
\emph{Fract. Calc. Appl. Anal.}  {\bf 7}, No 3 (2004), 339-364.

\bibitem{luchko_etal-fcaa-2007}
Yu. Luchko, J. Trujillo,  
Caputo-type modification of the Erd\'elyi-Kober fractional derivative.  
\emph{Fract. Calc. Appl. Anal.} {\bf 10}, No 3 (2007), 249--267.

\bibitem{lundstrom_etal-nn-2008}
B.N. Lundstrom, M.H. Higgs, W.J. Spain, A.L. Fairhall,
Fractional differentiation by neocortical pyramidal neurons.
{\it Nature Neuroscience} {\bf 11}, No 11 (2008), 1335--1342.

\bibitem{mainardi-csf-1996}
F. Mainardi,
Fractional relaxation-oscillation and fractional diffusion-wave phenomena.
{\it Chaos, Solitons \& Fractals} {\bf 7}, No 9 (1996), 1461--1477.

\bibitem{mainardi-2010}
F. Mainardi,
{\it Fractional Calculus and Waves in Linear Viscoelasticity}.
Imperial College Press, London (2010).

\bibitem{mainardi_etal-fcaa-2001}
F. Mainardi, Y. Luchko, G. Pagnini, 
The fundamental solution of the space-time fractional diffusion equation. 
\emph{Fract. Calc. Appl. Anal.} \textbf{4}, No 2 (2001), 153--192.

\bibitem{mainardi_etal-ijde-2010}
F. Mainardi, A. Mura, G. Pagnini,
The $M$-Wright function in time-fractional diffusion processes: A tutorial survey.
{\it Int. J. Diff. Equations} {\bf 2010}, (2010), 104505.

\bibitem{mainardi_etal-jcam-2007}
F. Mainardi, G. Pagnini,
The role of the Fox--Wright functions in fractional sub-diffusion of distributed order. 
{\it J. Comput. Appl. Math.} {\bf 207}, No 2 (2007), 245–-257.

\bibitem{mainardi_etal-fcaa-2003}
F. Mainardi, G. Pagnini, R. Gorenflo,
Mellin transform and subordination laws in fractional diffusion processes.
\emph{Fract. Calc. Appl. Anal.} \textbf{6}, No 4 (2003), 441--459.

\bibitem{metzler_etal-jpa-2004}
R. Metzler, J. Klafter, 
The restaurant at the end of the random walk: recent developments in fractional dynamics descriptions of anomalous dynamical processes.
{\it J. Phys. A: Math. Gen.} {\bf 37}, No 31 (2004), R161--R208.

\bibitem{mura-phd-2008}
A. Mura,
{\it Non-Markovian Stochastic Processes and Their Applications: From Anomalous Diffusion to Time Series Analysis}.
Ph.D. Thesis, University of Bologna (2008).
http://amsdottorato.cib.unibo.it/846/1/Tesi\_Mura\_Antonio.pdf\\
Now available by Lambert Academic Publishing (2011).

\bibitem{mura_etal-itsf-2009}
A. Mura, F. Mainardi,
A class of self-similar stochastic processes with stationary increments to model anomalous diffusion in physics.
{\it Integr. Transf. Spec. Funct.} {\bf 20}, No 3 (2009), 185--198.

\bibitem{mura_etal-jpa-2008}
A. Mura, G. Pagnini,
Characterizations and simulations of a class of stochastic processes to model anomalous diffusion.
{\it J. Phys. A: Math. Theor.} {\bf 41}, No 28 (2008), 285003. 

\bibitem{mura_etal-pa-2008}
A. Mura, M.S. Taqqu, F. Mainardi, 
Non-Markovian diffusion equations and processes: Analysis and simulations.
{\it Physica A} {\bf 387}, No 21 (2008), 5033--5064.

\bibitem{pagnini-fcaa-2011}
G. Pagnini,
Nonlinear time-fractional differential equations in combustion science.
\emph{Fract. Calc. Appl. Anal.} \textbf{14}, No 1 (2011), 80--93.

\bibitem{pagnini-epjst-2011}
G. Pagnini,
The evolution equation for the radius of a premixed flame ball in fractional diffusive media.
{\it Eur. Phys. J. Special Topics} {\bf 193}, (2011), 105--117.

\bibitem{podlubny-1999}
I. Podlubny,
{\it Fractional Differential Equations}. 
Academic Press, San Diego (1999).

\bibitem{rocco_etal-pa-1999}
A. Rocco, B.J. West, 
Fractional calculus and the evolution of fractal phenomena. 
{\it Physica A} {\bf 265}, No 3-4 (1999), 535--546.

\bibitem{saxena_etal-pa-2011}
R.K. Saxena, G. Pagnini,
Exact solutions of triple-order time-fractional differential equations for anomalous relaxation and diffusion I: 
The accelerating case. 
{\it Physica A} {\bf 390}, No 4 (2011), 602–-613.

\bibitem{scalas-pa-2006}
E. Scalas,
The application of continuous-time random walks in finance and economics.
{\it Physica A} {\bf 362}, No 2 (2006), 225--239.

\bibitem{schneider-1990} 
W.R. Schneider, 
Grey noise.
In: {\it Stochastic Processes, Physics and Geometry}, 
S. Albeverio, G. Casati, U. Cattaneo, D. Merlini, and R. Moresi (Editors), 
World Scientific, Teaneck (1990), 676--681. 

\bibitem{schneider-1992}
W.R. Schneider, 
Grey noise.
In: {\it Ideas and Methods in Mathematical Analysis, Stochastics, and Applications} vol. I,
S. Albeverio, J.E. Fenstad, H. Holden and T. Lindstr{\o}m (Editors),
Cambridge University Press, Cambridge (1992), 261--282.

\bibitem{sneddon-1966}
I.N. Sneddon, \emph{Mixed Boundary Value Problems in Potential Theory}. 
North-Holland Publ., Amsterdam (1966).

\bibitem{sneddon-lnm-1975} 
I.N. Sneddon,
The use in mathematical analysis of the Erd\'elyi--Kober operators and some of their applications, 
In: \emph{Lect. Notes Math.} {\bf 457}, Springer--Verlag, New York (1975), 37--79.

\bibitem{sneddon-1979}
I.N. Sneddon, 
\emph{The Use of Operators of Fractional Integration in Applied Mathematics}. 
RWN -- Polish Sci. Publ., Warszawa-Poznan (1979).

\bibitem{sokolov_etal-appb-2004}
I.M. Sokolov, A.V. Chechkin, J. Klafter, 
Distributed-order fractional kinetics. 
{\it Acta Phys. Pol. B} {\bf 35}, No 4 (2004), 1323–-1341.

\bibitem{tenreiro-machado-fcaa-2011}
J.A. Tenreiro Machado,
And I say to myself: ``What a fractional world!''.
\emph{Fract. Calc. Appl. Anal.} \textbf{14}, No 4 (2011), 635--654.

\bibitem{vinagre_etal-fcaa-2000}
B.M. Vinagre, I. Podlubny, A. Hern\'andez, V. Feliu, 
Some approximations of fractional order operators used in control theory and applications.
\emph{Fract. Calc. Appl. Anal.} \textbf{3}, No 3 (2000), 231--248.

\end{thebibliography}
\end{document}